# Fixed point of second virial coefficients in the glass transition

Jialin Wu

College of Material Science and Engineering, Donghua University,
Shanghai, 200051, China, E-mail: jlwu@dhu.edu.cn

**Abstract** Classical thermodynamic theory still holds true in subsystem that is a percolation connected by 8 orders of self-similar 2-body-3-body coupling clusters. The fixed point, $B_2^* \equiv 3/8$, for the clusters of different size, existing in reduced second Virial coefficients has been proved by scaling theory in percolation field. It is shown that, if $B_2^* \equiv 3/8$ is combined with $B_3^* \equiv 5/8$, the potentials of 2-body-3-body coupling clusters, in critical local cluster growth phase transition, balance the kinetic energy in the glass transition. It is also proved that the glass transition corresponds to the regime in which the chemical potentials in all subsystems hold zero.

## I INTRODUCTION

The confluence of both the thermodynamic and the kinetic dimensions of the liquid↔glass transition has presented one of the most formidable problems in condensed matter physics [1, 2]. In the previous paper [3], a theoretical framework of the intrinsic 8 orders of 2-*D* mosaic geometric structures has been proposed, from which many currently-prevalent theories and explanation on the glass transition can be involuntarily deduced. Furthermore, the paper of Fixed point for the Lennard-Jones potentials in the glass transition [4] also has provided an independent testification that there are only 8 orders of self-similar hard-spheres in the glass transition. The paper [4] moreover has proved that (a) the origin and transfer of interface excitation in the glass transition come of the balance effect between self-similar Lennard-Jones (L-J) potential fluctuation and geometric phase potential fluctuation; (b) a universal behavior: two orthogonal degenerate states, the fast reduced geometric phase factor 3/8 and the slow reduced geometric phase factor 5/8, is accompanied with the appearance of each order 2-*D* *symmetrical interface excitation closed loop* (SIECL) in the glass transition. The sum of the two degenerate reduced phase factors is exactly equal to 1. The geometric phase factor here is exactly the non-integrable phase potential to induce cluster migration. Thereby, the further cognition to the relationship between the random system and the glass transition can be obtained: in the thermal random motion system during glass thaw, there are at least five kinds of ordered motions (potential energy) which are balanced with disordered motions (kinetic energy).

The first motion is fluctuation self-similarity. The self-similarity potential fluctuations of local and global, specially selected from random potentials fluctuations, leads to the lower attractive potential of $-17/16\varepsilon_0$ ($\varepsilon_0$, potential well energy) and the interface excitation and excited-interface transfer in the glass transition [4].

The second motion is local 2-*D* SIECL and localized energy. Some excited interfaces, selected from random excited interfaces, form the instantaneous *i*-th order 2-*D* SIECL in local z-component space (or on x-y projection plane), which defines *i*-th order cluster. In random motion, once a 2-*D* SIECL forms, the surrounded cluster will gain non-integrable phase potential to induce cluster migration on the z-axial. According to the topological analysis, the cluster is always in slow movement, the $2\pi$ energy flow is always in comparatively fast movement and the former is induced by $2\pi$ energy flow of SIECL on the cluster interfaces. The localized energy independent of temperature, $E_c = kT_g^* = 20/3\varepsilon_0$ [3], shows that in the ideal random motion, the symmetrical 2-*D* closes loop is neither optional nor arbitrary large. In the temperature range from $T_g$ to $T_m$, there are only 8 orders of symmetrical 2-*D* closes loops with 8 orders of relaxation times in the ideal random motion. These loops form inverse energy cascade from small to large, whose energy is localized energy $E_c$. The largest 2-*D* SIECL is the 8th order 2-*D* SIECL mentioned in [3].

The third motion is cooperative orient migration in local field, in which the 8 orders of 2-*D* SIECLs appear from small to large, from fast to slowly, in the manner of inverse cascade. This is also the manner of breaking a solid domain. When the 8th order loop completes, its energy will transfer to random vibration kinetic energy in form of fast cascade vibration and the re-built excited interfaces will begin new inverse cascade.

The fourth motion is the degree of freedom of flow-percolation. The increase of system temperature nothing but increase the number of inverse cascade-cascade in local zone, it can not increase the size and energy of 2-*D* SIECL. That is just the physical meaning of intrinsic localized energy in the thermo random system. The glass transition corresponds to the critical transition in which all the local inverse cascade motions are just connected (i.e. flow-percolation, which is the sub-system in system). At that time, each *i*-th order loop gets the evolution energy $\varepsilon_0$ of one outer degree of freedom of the *i*-th order cluster and the free motion energy $5\varepsilon_0$ of five inner degree of freedom of the centric (*i*-1)-th order cluster in the *i*-th order loop. Thus, the degree of freedom of flow-percolation in the glass transition is 1 [3]. The melt transition means that the outer degree of freedom of flow-percolation is 5.

The fifth motion is dynamical mosaic structure of flow-percolation [3], which makes the evolvement of each reference local field consist of delaying contribution of 4 neighboring local fields to the evolvement of central local field. Thereupon, each excited domain does not return to the solid state for the disappearance of excited interfaces in the initial reference local field, instead, they always consist of excited interfaces of delaying contribution to the excited domain by the 4 local fields neighboring update reference local field.

Thermo disorder-induced localization is also one of the most important properties from $T_g$ to $T_m$ in the solid-liquid transition. In the paper [3], the localized energy $E_c = 20/3\varepsilon_0$ in the glass transition has been deduced by geometry method and in this paper, $E_c \approx 20/3\varepsilon_0$ will be directly proved by statistic thermodynamics.

On the other way, although the instant 8 orders of 2-*D* mosaic geometric structures have been found by geometry method, all the other loops will not appear as long as the 8th order loop has not appeared, as emphasized in the paper [3]. The problem is that in the critical state of the 8th order loop being about to appear, how the 8 orders of SIECL in the system realize it by the most effective way and that since the 8 orders of loops are the saltative loops, how the fluctuation physical picture of the system is described at the most local time. The two questions above will be answered by the geometry method mentioned in [3], namely, the standout mode of an arrow on an excited interface, in one solution. In the critical 8 orders of local cluster growth phase transitions (LCGPP) of glass transition, the fluctuating 8 orders of *self-similar 2-body-3-body coupling clusters* have appeared and their fluctuation interfaces will one by one abrupt change in the most effective way in order to realize 8 orders of 2-*D* mosaic geometric structures and complete the glass transition. In other words, there are 8 orders of self-similar and fluctuating 2-body-3-body coupling clusters that excited by thermo random motion in the glass transition.

If this idea is correct, it will be expected that the sum of two potential energies of 2-body and 3-body should balance with the thermal random motion energy $kT$ (current hard sphere model can not do as this [5]). This will be the sixth motion that potential balances kinetic energy. This motion will be discussed by way of the reduced second, third Virial coefficients in this paper. In the percolation field, all the excited locals are connected as a whole. While in the critical state, each excited local can also be seen as 8 orders of self-similar 2-body-3-body clusters in the fluctuation-evolution states. Therefore, in a percolation field, all the $i$-th order 2-body-3-body clusters are connected with each other, namely, the $i$-th order inverse cascade energy flow in the percolation field, named as $V_i$-percolation field and $V_i$ is the volume of the $i$-th flow. The second and the third Virial coefficients for $i$-th order 2-body and 3-body will be discussed respectively.

This paper mainly aims to prove that the second Virial coefficients of clusters of different size exist fixed point, by using the classical thermodynamic theory and the scaling theory. It will be further proved that the potentials of 2-body-3-body coupling clusters balance the kinetic energy in the glass transition.

## II THEOREM PROVE FOR $B_2^* \equiv 3/8$

### A. Two-body-three-body coupling clusters in critical local cluster growth phase transition

In the LCGPP of glass transition, if each 2-$D$ SIECL appears in the way of one arrow after another according to the appearance probability as mentioned in figure 3-6 in the paper [3], the probability is rather low. In fact, the more probable situation is that each loop in the 8 2-$D$ SIECLs first form two fluctuation 3-body clusters in the space and when the 8th loop appears, 8 orders of 2-body-3-body coupling clusters cooperatively reduce the total excited energy to, one by one, form 8 order 2-$D$ SIECLs and eliminate interfaces in the loop. As shown in Fig.1 (*a*), in a reference $a_0$ particle $V_0(a_0)$, at different times and in different spaces and positions, respectively form two 3-body clusters: $V_0(a_0)$ +$V_0(b_0)$ + $V_0(c_0)$ and $V_0(a_0)$ + $V_0(d_0)$ +$V_0(e_0)$, with neighboring particle $V_0(b_0)$, $V_0(c_0)$ and particle $V_0(d_0)$, $V_0(e_0)$. Each 3-body cluster only needs the excitation energy of 10 excited interfaces, which is less than that of 12 excited interfaces in Fig.1 (*b*). The two particle $V_0(a_0)$ in two 3-body clusters, excited at different times, move randomly and they are not in superposition. Once they are in superposition, Fig. (*a*) immediately mutates into Fig. (*b*) to form the first order SIECL and the excitation interfaces inside the loop disappear. In the critical state of the 8th loop's appearance, 8 orders of clusters of local field appear in the form of two fluctuation 3-body clusters as shown in Fig. 1 (*a*), named as 2-body-3-body coupling clusters in the glass transition.

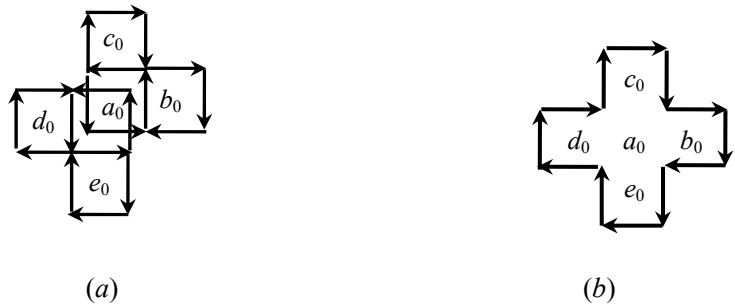

(*a*)  (*b*)

**Figure1.** Schematic diagram of 2-body-3-body coupling cluster in the critical LCGPP of glass transition. When the two centers of $V_0(a_0)$ of two fluctuation three-body clusters, $V_0(a_0)$+$V_0(b_0)$ +$V_0(c_0)$ and $V_0(a_0)$+$V_0(d_0)$ +$V_0(e_0)$, coincide, first order SIECL is formed (b).

### B. Virial expansion in the glass transition

The virial expansion is generally used to study the many-body problem. Eq. (1) is the expression of Virial expansion in classical gas system.

$$\frac{PV}{kT} = 1 + \frac{1}{b_0}B_2^0 + \frac{1}{b_0^2}B_3^0 + \cdots = 1 + B_2 + B_3 + \cdots \qquad (1)$$

Where $B_2 = B_2^0/b_0$ and $B_3 = B_3^0/b_0^2$ are the reduced second and third Virial coefficient respectively. The numerical value '1' in eq (1) comes from the contribution of repulsive potential to surface transformed from the kinetic energy of one gas quasi-particle itself. The physical picture in classical gas system is described as follows. One quasi-particle *moves* along *one direction* and its motion is instantly *arrested* by surface in collision. The kinetic energy transforms into the potential. All other items in eq (1) are contributions of the kinetic energies from 2-body, 3-body and so on, transforming into $PV/kT$. $P$ in eq (1) is a positive pressure. The sign in front of particle volume is positive, which represents that the *outward kinetic energy contributing to interface repulsive potential* in classical thermodynamic theory.

The theoretical picture of ideal glass transition is exactly reversed to that of classical ideal gas. 2-body-3-body clusters (from 'static' to 'moving' along one direction to thaw a solid domain) should 'catch' outside kinetic energy to be excited and have 'attraction' potential. A reference $i$-th order cluster (hard-sphere) is defined by the $i$-th order SIECL contributed by its 4 neighboring ($i$-1)-th order loop energies [3] and the volume sign of $i$-th order cluster is the same with that of $i$-th order loop. Here, one of the key proofs is that the direction of the $i$-th order loop (the direction along which the $i$-th order cluster is driven to move) is always opposite to that of the ($i$+1)-th loop, in other words, in critical LCGPP, the *free motion*, being of 5 degrees of freedom, of $i$-th order cluster always appears in the ($i$+1)-th induced potential field whose sign of induced direction is opposite to that of the $i$-th cluster (this will give rise to the jamming effect [3]). Thus, in order to refract 'the *phase difference of $\pi$* between the cluster in free motion state and the induced potential', (i.e., the phase difference of $\pi$ between kinetic energy and potential energy, which is also one of the singularities of glass transition), a 'negative sign' should be attached to the front of the hard-sphere volume in the picture of critical LCGPP. That is, glass transition is a phenomenon of transformation from *inward kinetic energy contributing to interface attraction potential*. The role reversal between 'outward kinetic-repulsive potential' and 'inward kinetic- attraction potential' in physics shows that the classical thermodynamic theory still holds true for glass transition in percolation field. In eq (1), the contribution of item '1' to $PV/kT$ is replaced by that of a 2-body-3-body coupling cluster. This is because that the interface of hard-spheres is formed by interface excitation and each interface excitation always connects with two clusters. Therefore, there is no *single* cluster whose attraction potential alone contributes to $PV/kT$. Any items after the third item in eq (1) are also replaced by self-similar 2-body-3-body coupling clusters, which do not conflict with the mode-coupling theory [6]. In eq (1) the positive pressure is changed to negative attraction tensile stress which keeps the balance to the exterior stress on percolation field in the glass transition. In order to conveniently apply the classical thermodynamic, a 'negative sign' should be respectively attached to the front of the stress $P$ and the volume $V$ in this paper. The sign of $PV/kT$ is still positive. Eq (2) is a Virial expansion around the volume variable $V$ of connected clusters on percolation field in the glass transition, where $V$ is only the *volume of sub-system* (percolation field), instead of the volume of the whole system.

$PV/kT = B_2 + B_3$ \qquad (2)

**C. Self-similar equations of two-body clusters in the glass transition**

From Enthalpy $H = E + PV$, the definition of Joule-Thomson coefficient $\mu_J$ [7] is

$$\mu_J = (\partial T / \partial P)_{H,N} = C_P^{-1}\left[T(\partial V / \partial T)_{P,N} - V\right] \qquad (3)$$

Assume the state of $\mu_J \equiv 0$ corresponds to glass transition. When $\mu_J \equiv 0$

$$(\partial V/\partial T)_{P,N} \equiv V/T \tag{4}$$

This is the volume change of sub-system as a function of temperature when outside pressure (stress) remains constant. As long as the condition $\mu_J \equiv 0$ in eq (3) is satisfying, eq (4) holds true. In addition, we see, $C_P$ in eq (3) may show an *abnormal change* during glass transition. It can also be strictly proved that the reduced third Virial coefficient for hard-sphere system is *constant 5/8*; independent of temperature [8], i.e., $B_3$ in eq (2) is independent of temperature. The substitution of eq (4) into eq (2) yields

$$\frac{d(B_2/T)}{dT} \equiv 0 \quad \text{or} \quad \frac{dB_2}{dT} \equiv \frac{B_2}{T} \tag{5}$$

Eq (4) and (5) show that the physical quantities of $dB_2/dT$ and $\partial V/\partial T$ on small scale are respectively equal to those of $B_2/T$ and $V/T$ on large scale, which represents the self-similarity between the small scale and the large scale in critical phase transition. Eq (4) and (5) thus are the self-similar phase transition equations of *sub-systems* in glass transition. Note that the enthalpy $H$ in eq (3) is *invariant* in any sub-systems but not in the whole system. These self-similar phase transition equations of sub-systems, together with $df/d\chi \equiv f/\chi$ mentioned in previous paper [4], are in fact the condition equations for occurrence of self-similar 2-body-3-body clusters in the glass transition.

### D. Approximate solution for self-similar equations

In the previous paper [4], one of the important results is that the definition or origin of hard-sphere can be also deduced from the self-similar Lennard-Jones (L-J) potentials in ideal random system. Here the key point is the self-similar in system. Now the approximate solution is deduced from the self-similar eq (5) also by L-J potential. Take [9]

$$B_2(T^*)_{L-J} \equiv B_2^0(T^*)_{L-J}/b_0 = \frac{4}{T^*}\int_0^\infty dx \cdot x^2 \left[\frac{12}{x^{12}} - \frac{6}{x^6}\right] \exp\left\{-\frac{4}{T^*}\left[\left(\frac{1}{x}\right)^{12} - \left(\frac{1}{x}\right)^6\right]\right\} = \sum_{n=0}^{\infty} \alpha_n \left(\frac{1}{T^*}\right)^{\frac{2n+1}{4}} \tag{6}$$

In this equation $\alpha_n$ can be represented by the $\Gamma$ function: $\alpha_n = -\frac{\sqrt{2}\Gamma\left(\frac{2n-1}{4}\right)}{2^{(2-n)}n!}$ in Eq. (6), $x = q/\sigma$. $q$, $\sigma$ are all parameters in the L-J potential. $T^* = kT/\varepsilon_0$ is the reduced temperature, $\varepsilon_0$, the L-J potential-well energy.

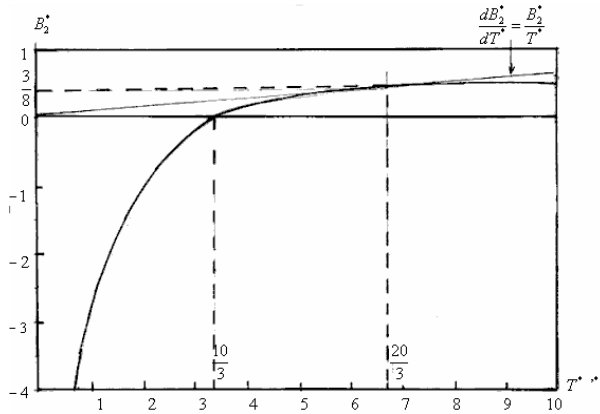

**Figure2.** Diagrammatize to equations (5) and (6). In the figure, $T_g^* \approx 20/3$ denotes the reduced localized random kinetic energy point during glass transition when $B_2^*(20/3) \approx 3/8$. $T_g^*$ is a fixed point independent of temperature.

There is no analytical solution after substituting eq (6) into eq (5). Graphical method gives the only set of approximate solution, see Fig.1.

$$\begin{cases} T_g^* \approx 20/3 \\ B_2^*(T_g^*) \approx 3/8 \end{cases} \tag{7}$$

The numerical solution in Fig. 2 refers to the result of [9]. There are two temperature points on the curve of $B_2(T)$, respectively corresponding to the two special *random motion energy* and denoted as $T_g^* \approx 20/3$ and $T_0^* > 10/3$. Only the point $T_g^*$ will be discussed in this paper. Fig.2 shows that in the common expression of the second virial coefficient in the ideal random system, only the point $(T_g^*, B_2^*)$, which corresponds to the LCGPP in the glass transition, meets the condition of self-similarity. The "temperature" point $T_g^*$ seemingly corresponds to that of the glass transition, but does not generally do unless the 'slow process' nondimensional heat motion energy $kT/\varepsilon_0$ supplied from environment is exactly equal to $T_g^*$. Due to the fact that eqs (2) – (7) only come into existence on sub-system, $T_g^*$ has no option but to be the LCGPP temperature in sub-system, that is, $T_g^*$ is the *reduced critical random energy* in sub-system (flow-percolation) for the occurrence of 8 orders 2-body-3-body cluster*s*. $T_g^*$ is called as the *intrinsic* and *invariant* random kinetic energy which is an important physical quantity to characterize the intrinsic property in the glass transition. According to Anderson [1] theory of disorder-induced localization, $T_g^*$ is defined as the (nondimensional) *localized energy* of thermo random motion (kinetic energy) induced by thermo-disorder in glass transition. From Fig.1, the localized energy, $E_c = T_g^*\varepsilon_0 \approx 20/3\ \varepsilon_0$, approximately equals to the value deduced from the previous paper [3].

There are 8 orders of localized energies $T_g^*\varepsilon_0(\tau_i)$ in glass transition, in which $\tau_i$ denotes the relaxation time of *i*-th order hard-sphere. The fixed point $T_g^*$ in Fig.1 shows that the numerical value of $T_g^*$ is invariable, but it contributes to different order of clusters in inverse energy cascade. Eqs (2) to (6) will always be valid and $B_2^*$ will always equal to 3/8, regardless if the temperature $T_g$ ($T_g$ is the apparent glass transition temperature) will further increase. Point $(T_g^*, B_2^*) = (20/3, 3/8)$ is a stable fixed point. *When $T_g$ further increases, the increase of randomness in system cannot be otherwise than the increase of the number* of *inverse cascade- clusters.*

This foreshows that, on the fixed point of $T_g^*$ in the glass transition, there is an identical second Virial coefficient $B_2^*(T_g^*)$ for all difference clusters in size. Thus, by first determining the general second Virial coefficient expression $B_2(V_i)$, which depends on the order number, *i*, of clusters, and then the scaling equation for different clusters in size, $B_2^*$ can be finally deduced from the fixed point of the scaling equation.

Diagrammatic of Fig. 1 indicates：$B_2^*(T_g^*) \approx B_2^*(20/3) \approx 3/8$

### E. Fixed point of second Virial coefficient: theorem prove for $B_2^* \equiv 3/8$

$B_2^* \approx 3/8$ in eq. (7) is an approximate solution using L-J potential. Now $B_2^* \equiv 3/8$ is directly proved by scaling theoretical approach. The first step of testification is to find out the physical quantity- Fugacity, which can reflect the contribution of clusters of different size to Virial coefficient expression in the classical Virial coefficient expression. Following the derivation for $B_2$ in classical statistical physics see as in [9]

Pressure (stress)

$$p = -\frac{\Omega(V,T,\mu)}{V} = kT \sum_{l=1}^{\infty} \frac{b_l(T,V)e^{\beta l \mu}}{\lambda^{3l}} \tag{8}$$

Particle density

$$\frac{\langle N \rangle}{V} = -\frac{1}{V}\left(\frac{\partial \Omega}{\partial \mu}\right)_{V,T} = \sum_{l=1}^{\infty} \frac{l b_l(T,V) e^{\beta l \mu}}{\lambda^{3l}} \qquad (9)$$

Power series of Virial expansion

$$\frac{pV}{\langle N \rangle kT} = \sum_{l=1}^{\infty} B_l(T)\left(\frac{\langle N \rangle}{V}\right)^{l-1} \qquad (10)$$

The symbols are the same as in [9], where $\beta = 1/kT$, $\mu$: the chemical potential, $\beta_l$: the $l$ order of Virial coefficient, $b_l$: the $l$ – cluster integral. Eqs (8) – (10) hold true for any sub-system in glass transition. Take the thermodynamic limits for all equations of (8) – (10), i.e., $V \to \infty$, and notice that the limit condition corresponds to percolation field, which allows strange shapes [10] and here a fractal shape.

$<N> \to \infty$. $<N>/V$ is constant. $b_l(T,V) \to \overline{b}_l$. Combine (9) – (10), and yield:

$$\left(\sum_{l=1}^{\infty} \frac{\overline{b}_l(T) e^{\beta l \mu}}{\lambda^{3l}}\right)\left(\sum_{n=1}^{\infty} \frac{n \overline{b}_n(T) e^{\beta n \mu}}{\lambda^{3n}}\right)^{-1} = \sum_{l'=1}^{\infty} B_{l'}\left(\sum_{n'=1}^{\infty} \frac{n' \overline{b}_{n'}(T) e^{\beta n' \mu}}{\lambda^{3n'}}\right)^{l'-1} \qquad (11)$$

Expand both sides of eq. (11) and make the two second power coefficients of $\lambda^{-3} e^{\beta \mu}$ equal, and the second Virial coefficient is obtained. Notice that the *factors* of $e^{\beta \mu}$ on both sides of eq (11) *can be canceled out*.

$$B_2(T) = -\overline{b}_2(T) \qquad (12)$$

Eq (12) is the second Virial coefficient expression of two-particle cluster for a general system. As is stated in the introduction, it is necessary to respectively discuss the second Virial coefficient for the two-body clusters of different size at $T_g$ temperature in the critical LCGPP. It is also necessary, thereby, to modify eq (10) and (2) to the $i$-th order cluster with cluster volume $v_i$ on $V_i$-percolation field

$$(PV/kT_g)_i = B_2(v_i) + B_3(v_i) \qquad (13)$$

$$B_2(v_i) = \frac{(PV)_i}{kT_g} \quad (i = 1, 2, \ldots 8) \qquad (14)$$

$$B_3(v_i) \equiv 5/8 \qquad (15)$$

In the eq (13) and (14), the subscript $i$ represents $V_i$-percolation field. Eq (14) denotes the contribution from the $i$-th order two-body interaction to $PV/kT$ on $V_i$-percolation field. It should be noted that the result of eq (15) is still correct, though it is the referenced result from [8], while the testification of [8] is just according to the two preconditions of hard-sphere with positive repulsive force and the random distribution of distance between hard-spheres. The hard-sphere with negative attraction potential and two-body interaction potential in glass transition just depend on the distance-increment between two-body, while the random distribution of distance-increment can be obtained from two random distributions of distance between hard-spheres and eq (15) can be obtained from each of the latter.

Since in general gas-state system, only the second Virial coefficient for two-body cluster of a certain size is considered, an important physical quantity – Fugacity, $e^{\beta \mu} \equiv K$, is canceled out in the deduction of eq (12) from eq (11). There are 8 orders of different (relaxation time and size) two-body clusters in glass transition. Fugacity takes a key role in scaling transformation. From eq (8) it is found that $P/kT$ has the factor of $e^{\beta \mu}$, therefore, eq (12) may be rewritten as

$$B_2(v_i, K_i) = -\overline{b}_2(v_i) e^{\beta \mu_i} = -\overline{b}_2(v_i) K_i \qquad (16)$$

Note that $V \to \infty$ as eq (12) is derived. This corresponds to the condition of $i$-th order inverse cascade energy flow in a percolation field. Percolation is a phase transition of geometric connection at a certain probability. Due to fluctuation effects, the variation of $K_i$, arose from the number change of the $i$-th order clusters in $V_i$-percolation field, is denoted as $\Delta K_i$ and in a similar way, the variation of $K_{i+1}$ is denoted as $\Delta K_{i+1}$. From eq (16)

$$\begin{cases} \Delta B_2(v_i, \Delta K_i) = -\overline{b_2}(v_i)\Delta K_i \\ \Delta B_2(v_{i+1}, \Delta K_{i+1}) = -\overline{b_2}(v_{i+1})\Delta K_{i+1} \end{cases} \quad (17)$$

At $T_g$, from eq (14), eq (17) may be rewritten as

$$\frac{\overline{b_2}(v_i)\Delta K_i}{\overline{b_2}(v_{i+1})\Delta K_{i+1}} = \frac{\Delta(PV)_i}{\Delta(PV)_{i+1}} \quad (18)$$

In [4], $(PV)_{i+1}$ represents the migration work made by the $(i+1)$-th order non-integrable phase potential when the $(i+1)$-th order SIECL formed by $i$-th order clusters finishes. Since inverse cascade does not dissipate energy, the energy of inverse cascade is invariable of $T_g*\varepsilon_0(v_i)$, while forming cluster costs interface excited energy and $\Delta(PV)_{i+1}$ cannot be otherwise than the energy of *number increase* of $i$-th order clusters, or, the energy of number increase of $i$-th order inverse cascade energy flow in percolation. The energy that comes of the number increase of inverse cascade of $i$-th order equals to the increment of random thermo-motion energy $\Delta(kT_g(v_i))$ of $v_i$ clusters in sub-system. And $\Delta(kT_g(v_i) - T_g*\varepsilon_0(v_i)) = k\Delta T_g(v_i)$. So, $\Delta(PV)_{i+1} = k\Delta T_g(v_i)$. From eq (5), for cluster volume $v_i$ at $T_g$, $B_2(v_i) = \Delta B_2(v_i) \cdot kT_g(v_i)/ k\Delta T_g(v_i)$, and from (14), $B_2(v_i)$ has the form

$B_2(v_i) = \frac{kT_g(v_i)}{k\Delta T_g(v_i)} \cdot \Delta B_2(v_i) = \frac{\Delta(PV)_i}{k\Delta T_g(v_i)}$, from eq (18), and $\Delta(PV)_{i+1} = k\Delta T_g(v_i)$, $B_2(v_i)$ takes the

form

$$B_2(v_i) \quad \frac{\overline{b_2}(v_i)\Delta K_i}{\overline{b_2}(v_{i+1})\Delta K_{i+1}} \quad (19)$$

If there exists fixed point of second Virial coefficient in eq (19), denoted as $B_2^*$, according to the definition of fixed point: $\overline{b}_2(v_i) = \overline{b}_2(v_{i+1})$, $B_2^*(v_i)$ is of the form

$$B_2^*(v_i) = \frac{\partial K_i}{\partial K_{i+1}} \quad (20)$$

The second step of testification is to deduce the $K_i$ scaling equation. It has been proved [4] that the ratio of the distance increment $\Delta q_{i+1}$ and the volume $\sigma_i$ between two $i$-th order clusters is constant, controlled by Lindermann ratio $\Delta q_{i+1}/\sigma_i = 0.1047$ and the direction of cluster growth is always along the local $q$-axial.

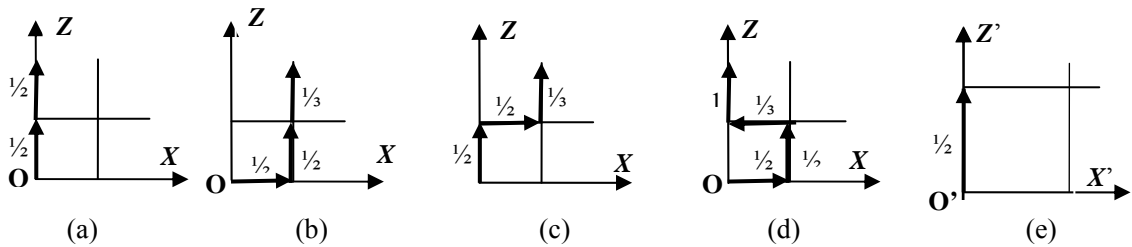

**Figure 3.** Schematic diagram of $K_i$ scaling transformation. Along z-axial, an $i$-th order cluster located at origin O has the 4 kinds of tracks (a), (b), (c), (d) evolving into an $(i+1)$-th order cluster (e) in thermo

random motion.

When the distance between the two $i$-th order clusters on $q$-axial is $q_i$ and the distance fluctuation is $\Delta q_{i+1}$, the two $i$-th order clusters are also the components of two $(i+1)$-th order clusters respectively. This picture should be equally represented by that of randomly distance fluctuation in Fig.3, in which an $i$-th order cluster on origin moves *randomly fluctuation* along a square with the side length of $\Delta q_{i+1}$ and the whole course is just the square on z-axial, i.e., the cluster randomly "walks out" the square alone + z-axial.

In Fig 3, the positive z-axial denotes the $q$-axial, which is also the increment direction of the cluster. The cluster can not evolve reversely midway, but evolve to the highest order [3] from small to large. That means the cluster can only move randomly along a side of square in the first quadrant. When it reaches the point of z =1 and continues to move along z-axial till outside the square, that means the cluster is moving in the $(i+1)$-th order cluster and is the component of it, or in other words, the $(i+1)$-th order cluster has formed. The $K_i$ scaling transformation depends on all (4 kinds of) possible tracks for a reference $i$-th order cluster (located at origin O) evolving into an $(i+1)$-th order cluster (along the positive z-axis, walking out the square). The 4 kinds of tracks are shown in Fig.3 (a-d). The weighting of $K_i$ is shown at the edge of each square in Fig. 3. The scaling equation of $K_i$ from Fig. 3 (a-d) is in the following form

$$\frac{1}{2}K_{i+1} = \left(\frac{1}{2}K_i\right)^2 + \left(\frac{1}{2}K_i\right)^2\left(\frac{1}{3}K_i\right) + \left(\frac{1}{2}K_i\right)^2\left(\frac{1}{3}K_i\right)K_i$$

$$= \frac{1}{4}K_i^2 + \frac{1}{6}K_i^3 + \frac{1}{12}K_i^4 \tag{21}$$

The 4 terms on the right side of eq (21) respectively represent the contributions from Fig.3 (a), (b), (c), (d) to (e) in scaling transformation. Denoting $K_c$ as the fixed point, from eq (21), the fixed point equation is

$$K_C = \frac{1}{2}K_C^2 + \frac{1}{3}K_C^3 + \frac{1}{6}K_C^4 \tag{22}$$

The solutions of the fixed point from eq (22) give: 0, 1, ∞. Because the critical point is always an instability fixed point, thus taking $K_c = 1$. From $K_c = 1 = e^{\beta\mu}$, we got the chemical potential in subsystem, $\mu \equiv 0$. Substituting eq (21) into eq (20) and taking $K_c = 1$, the result is

$$B_2^*(v_i) \equiv 3/8 \quad (i =1,2,\ldots 8) \tag{23}$$

This proof is self-consistent. If the macroscopic random heat energy $kT_g \geq T_g^*\varepsilon_0$, the increased random kinetic energy, $k\Delta T_g$, would increase the numbers of inverse cascade in system. If the mean random heat energy $kT$ satisfies: $1/8\varepsilon_0 < kT < T_g^*\varepsilon_0$, $1/8\varepsilon_0$ here is the interface excited energy [3, 4], due to the thermo fluctuation effects, it is possible for the local-excited energy-flows to form the geometric connected filed in a few of local fields within a long time, and low temperature glass transition thus occurs. *The condition for fluctuation stability is that the chemical potentials are always zero in all sub-systems.* Fugacity $K_i$ is a fixed point, $K_i \equiv 1$. Eq (12) is still derived from eq (16). The number of $i$-th order hard-spheres, $N_i$, in each order sub-system is invariant. Eqs (3) – (5) still hold true in all sub-systems. From eq (13), an impotent relationship equation has been deduced

$$(PV/kT_g)_i = B_2(v_i) + B_3(v_i) = 3/8 + 5/8 \equiv 1 \quad (i =1,2,\ldots 8) \tag{24}$$

Eq (24) holds true on all sub-systems, which means that the kinetic energy always keeps balances with the potential energy, in the manner of phase difference of $\pi$ (that is also one of the singularities in the glass transition), in the mode of 8 orders of 2-body-3-body clusters in the glass transition.

It should be noted that $B_2^* \equiv 3/8$ and $B_3^* \equiv 5/8$ intrinsically and respectively connect with the fast reduced geometric phase factor 3/8 and the slow reduced geometric phase factor 5/8 in the glass transition. In fact, any dimensionless distance increment (cavity volume) between two clusters all belongs to the component part of the reduced geometric phase factor and controlled by Lindemann ratio. The fast reduced geometric phase factor 3/8 is also the contribution of reduced random interactions of $i$-th order two-body cluster in ($i+1$)-th order loop, whereas the slow reduced geometric phase factor 5/8, refracted the $2\pi$ cyclic effect of $i$-th order two-body clusters, independent of cluster size, should also equals to the contribution of reduced three-body interactions of the minimum $2\pi$ closed loop.

### III. CONCLUSION

The percolation in glass transition is only a sub-system, which is connected by local fields in excited state. Classical thermodynamics still holds true in each sub-system, but not in the whole system. The conditions for the self-similar equations (3)-(5) in sub-systems permit heat capacity abnormality. From the solution of the self-similar equations we deduce the reduced localized energy of thermo random motion, $T_g^*$, which is a fixed point and independent of temperature, to characterize the universal intrinsic property in the temperature range from $T_g$ to $T_m$ in thermo random system. The value of $T_g^*$ is consistent with the theoretical value deduced from the previous paper [3] and validates the theory of the intrinsic 8 orders of 2-$D$ mosaic geometrical structures.

This paper proves that the reduced second Virial coefficients are constant, i.e., $B_2^* \equiv 3/8$ for self-similar two-body clusters. This paper also has proved that the glass transition occurs if the chemical potentials of sub-systems hold zero.

Here, it is an interesting and profound theoretical comparison that the chemical potential is defined Fermi energy on the occasion when the temperature is zero in solid physics; whereas the random motion energy, $kT_g$, corresponds to that of the glass transition on the occasion when the chemical potentials in sub-systems are always zero in solid-to-liquid transition.

President Yuan Tseh Lee and Professor Sheng Hsien Lin of the Academia Sinica, Taiwan supported the manuscript preparation. Valuable discussion with Professor Yun Huang of Beijing University, Professor Da-Cheng Wu of Sichuan University, Professor Bo-Ren Liang and Professor Cheng-Xun Wu of Donghua University are also acknowledged.